\documentclass{PoS}
\usepackage{physics}

\usepackage{setspace}
\usepackage[latin1]{inputenc}
\usepackage[T1]{fontenc}
\usepackage{soul}
\usepackage{graphicx}
\usepackage{lipsum,multicol}
\usepackage{cancel}
\usepackage{fancyhdr}
\usepackage{pgf,tikz}
\usetikzlibrary{arrows}
\usepackage{url}
\usepackage{chemfig}
\usepackage{braket}
\usepackage{mathrsfs}

\usepackage[bottom]{footmisc}
\usepackage{subfigure}
\usepackage[toc,page]{appendix}
\usepackage{gensymb}
\usepackage{mathrsfs}
\usepackage{sidecap}
\usepackage{wrapfig}
\usepackage{stmaryrd}
\usepackage{float}

{%
\stepcounter{equation}
\begin{equation*}}{%
\leqno(\arabic{equation})
\end{equation*}
}
\usepackage{xcolor}

\usepackage[customcolors,shade]{hf-tikz}

\title{Dark Matter searches towards the WLM dwarf irregular galaxy with H.E.S.S.}

\ShortTitle{DM searches towards WLM}

\author{Celine Armand\\
        Univ. Grenoble Alpes, Univ. Savoie Mont Blanc, CNRS, LAPP \& LAPTh, 74000 Annecy, France\\
        E-mail: \email{celine.armand@lapth.cnrs.fr}}
\author{Emmanuel Moulin\\
        IRFU, CEA, Universit\'e Paris-Saclay, F-91191 Gif-sur-Yvette, France\\
        E-mail: \email{emmanuel.moulin@cea.fr}}
        \author{Vincent Poireau\\
        Univ. Grenoble Alpes, Univ. Savoie Mont Blanc, CNRS, LAPP, 74000 Annecy, France \\
        E-mail: \email{poireau@lapp.in2p3.fr}}
\author{Lucia Rinchiuso\\
       IRFU, CEA, Universit\'e Paris-Saclay, F-91191 Gif-sur-Yvette, France\\
        E-mail: \email{lucia.rinchiuso@cea.fr}}

\author{for the H.E.S.S. collaboration\footnote{for collaboration list see PoS(ICRC2019)1177}}
\abstract{
In the indirect dark matter (DM) detection framework, the DM particles would produce some signals by self-annihilating and creating standard model products such as $\gamma$ rays, which might be detected by ground-based telescopes. 
Dwarf irregular galaxies represent promising targets for the search for DM as they are assumed to be dark matter dominated systems at all radii. 
These dwarf irregular galaxies are rotationally supported with relatively simple kinematics which lead to small uncertainties on their dark matter distribution profiles. 
In 2018, the H.E.S.S. telescopes observed the irregular dwarf galaxy Wolf-Lundmark-Melotte (WLM) for a live time of 19 hours. These observations are the very first ones made by an imaging atmospheric Cherenkov telescope toward this kind of object. We search for a DM signal looking for an excess of $\gamma$ rays over the background in the direction of the WLM galaxy. We present the first results obtained on the velocity weighted cross section for DM self-annihilation as a function of DM particle mass.}

\FullConference{36th International Cosmic Ray Conference -ICRC2019-\\
		July 24th - August 1st, 2019\\
		Madison, WI, U.S.A.}

\begin{document}

\section{Introduction}

Dark matter represents 85\% of all matter in the Universe, affecting the formation of large scale structures, influencing the motion of galaxies and clusters, and bending the path of light. Yet, we do not know much about its nature and properties. 


In the early Universe, dark matter particles such as the WIMPs (Weakly Interacting Massive Particles) are assumed to have been in full thermal equilibrium with the standard model (SM) particles at sufficiently high temperatures. Since the particle density is high, they can easily interact with one another. As the Universe expands, it gets less dense and cools down, which makes the interactions between particles less likely and the particle abundance freezes-out. Thus, dark matter particle annihilation is greatly suppressed but a relic density remains and dark matter particles still annihilate and may be observable in rich and dense regions such as dwarf galaxies or the Galactic center. Dark matter would then send some indirect signals by pair-annihilating and creating SM products, which might be detected. Among these particles used as probes for indirect dark matter searches are $\gamma$-rays. 
High-energy $\gamma$ rays offer several advantages: they are not deflected by the Galactic magnetic field, so that their source can be well localized in the sky. In addition, $\gamma$ rays do not undergo as much attenuation as the charged particles while propagating. This allows us to point directly our $\gamma$-ray telescopes to the sources to look for signals reaching the Earth.

The differential $\gamma$-ray flux (in $\gamma \cdot \text{m}^{-2} \cdot \text{s}^{-1} \cdot \text{GeV}^{-1}$) produced by dark matter annihilation in dwarf galaxies, assuming WIMPs are Majorana particles, is written as:

\begin{equation}
 \frac{d\Phi_{\gamma}}{dE} =\displaystyle{\frac{1}{2} \frac{\langle\sigma v\rangle}{4\pi m_{\chi}^2}} \: \frac{d\Phi_{PP}}{dE} \: \mathcal{J}
\label{big_formula}
\end{equation}


with $\displaystyle{\frac{d\Phi_{PP}}{dE}}$ given by
\begin{equation}
 \frac{d\Phi_{PP}}{dE} = \sum_f B_f  \frac{dN_{\gamma}^f}{dE_{\gamma}}  dE_{\gamma}
\end{equation}

and $\mathcal{J}$ by
\begin{equation}
  \mathcal{J} = \displaystyle{\int_{\Delta \Omega} \int_{\text{los}} \: \rho_{\text{DM}}^2(r(s, \alpha_{\text{int}})) ds d\Omega'  }.
\end{equation}



%

\vspace{0.3cm}

The first term is the normalization containing the DM mass $m_{\chi}$ and its annihilation cross section averaged over the velocity distribution $\langle\sigma v\rangle$. The second term is defined as the particle physics factor $d\Phi_{PP}/dE$ which encloses the differential spectrum $dN_{\gamma}^f/dE_{\gamma}$ of each annihilation channel $f$ ponderated by their branching ratio $B_f$. These differential spectra correspond to the number of $\gamma$ rays emitted per annihilation per energy range.
The last term is called the \textit{astrophysical $\mathscr{J}$ factor} describing the amount of dark matter annihilations occurring within the sources. This component holds the dark matter density profile $\rho_{\text{DM}}$ squared, as a function of the distance $r$ from the center of the galaxy. This squared density is then integrated along the line of sight (los) and over the solid angle $\Delta \Omega$. The solid angle corresponds to the field of view over which $\gamma$-ray telescopes (e.g. H.E.S.S.) observe the sky. 

This proceeding focuses on a new kind of target to probe dark matter: dwarf irregular galaxies.
Dwarf irregular galaxies (dIrrs) are very promising targets as they possess a $\mathscr{J}$ factor in the order of $\sim 10^{17}$ $\text{GeV}^2.\text{cm}^{-5}$. So far 36 of them have been optically observed within a distance of 11 Mpc and an extension of their halo of $0.3\degree<\theta_{\text{halo}}<3\degree$. These objects are rotationally supported with relatively simple kinematics. They are assumed to be dark matter dominated objects at all radii, even in their central part \cite{proceedingHAWC}. dIrrs offer the advantage to have well-constrained rotation curve which leads to an actual measured $\mathscr{J}$ factor (not a prediction) with very small uncertainties. Another property of these dwarf galaxies is their star-forming region, below 0.1\degree, at their center. The HAWC experiment published a study of irregular galaxies \cite{proceedingHAWC} and set limits on the DM annihilation cross section using these galaxies.
In 2018, the H.E.S.S. experiment observed one these dIrrs called WLM (Wolf-Lundmark-Melotte) which makes H.E.S.S. the first IACT (Imaging Air Cherenkov Telescopes) to observe this new kind of sources. 

H.E.S.S. is a Cherenkov telescope array located in central Namibia in the Khomas Highland plateau area, at around 1,800 meters above sea level. 
The original array consists of 4 small-sized telescopes (CT1-4) with 12-meter reflectors. Each of these reflectors is made of hundreds of spherical mirrors, concentrating the faint flashes on a camera installed in the focal plane of the telescope. These telescopes detect brief flashes of Cherenkov radiation generated by very high energy $\gamma$ rays of $\sim 100$ GeV up to $\sim 100$ TeV.
In 2012, a fifth, 28-meter telescope (CT5) was added to the array with an improved camera allowing detection at a lower threshold of $\sim 30$ GeV.  

\section{Properties of WLM}

WLM is a dwarf irregular galaxy located at (l = 75.86\degree, b = -73.62\degree) at 1 Mpc from the Milky Way. It possesses a star-forming region at its center and is isolated from other astrophysical sources. This dwarf possesses excellent HI data with a smooth HI distribution and a well-measured photometry and stellar kinematics \cite{R16} \cite{R18} with an extension of its halo of $r_{\text{halo}} = 49.4$ kpc ($\theta_{\text{halo}} = 2.89\degree$).
WLM is rotationally supported with no significant non-circular motions in the gas. A smooth rotation curve of this galaxy can then be derived, which is well-constrained from these measurements, and implies WLM is DM dominated \cite{R16}.

\section{DM distribution}

The DM distribution in WLM can be well represented by a coreNFW profile \cite{R18} that writes:
\begin{equation}
\rho_{\text{coreNFW}(r)} = \displaystyle{f^n(r) \rho_{\mathrm{NFW}}(r) + \frac{f^{n-1}(r) (1-f^2(r))}{4\pi r^2 r_c}M_{\mathrm{NFW}}(<r)}.
\end{equation}

\begin{figure}[H]
\centering{\includegraphics[scale=0.4]{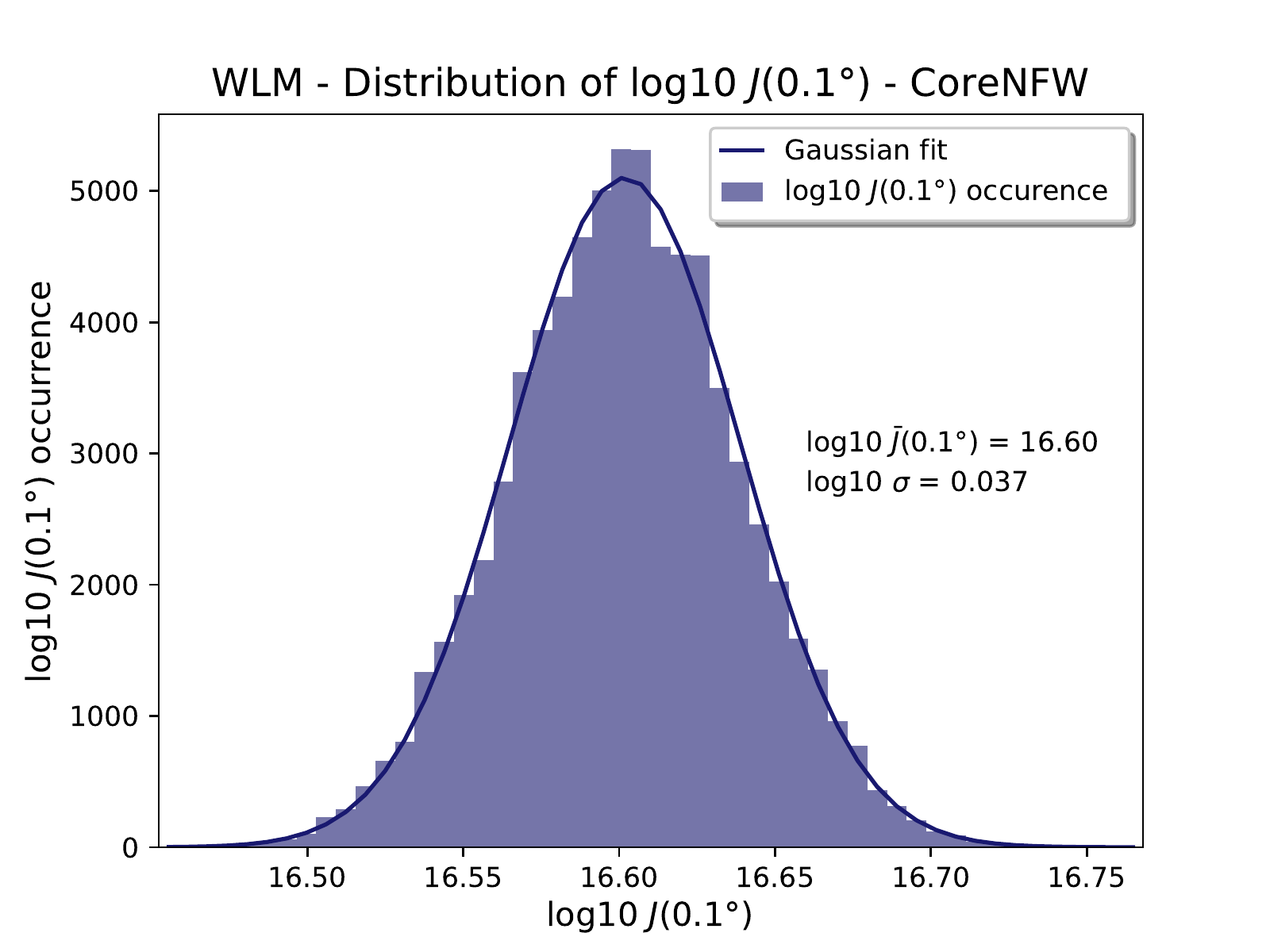} \includegraphics[scale=0.4]{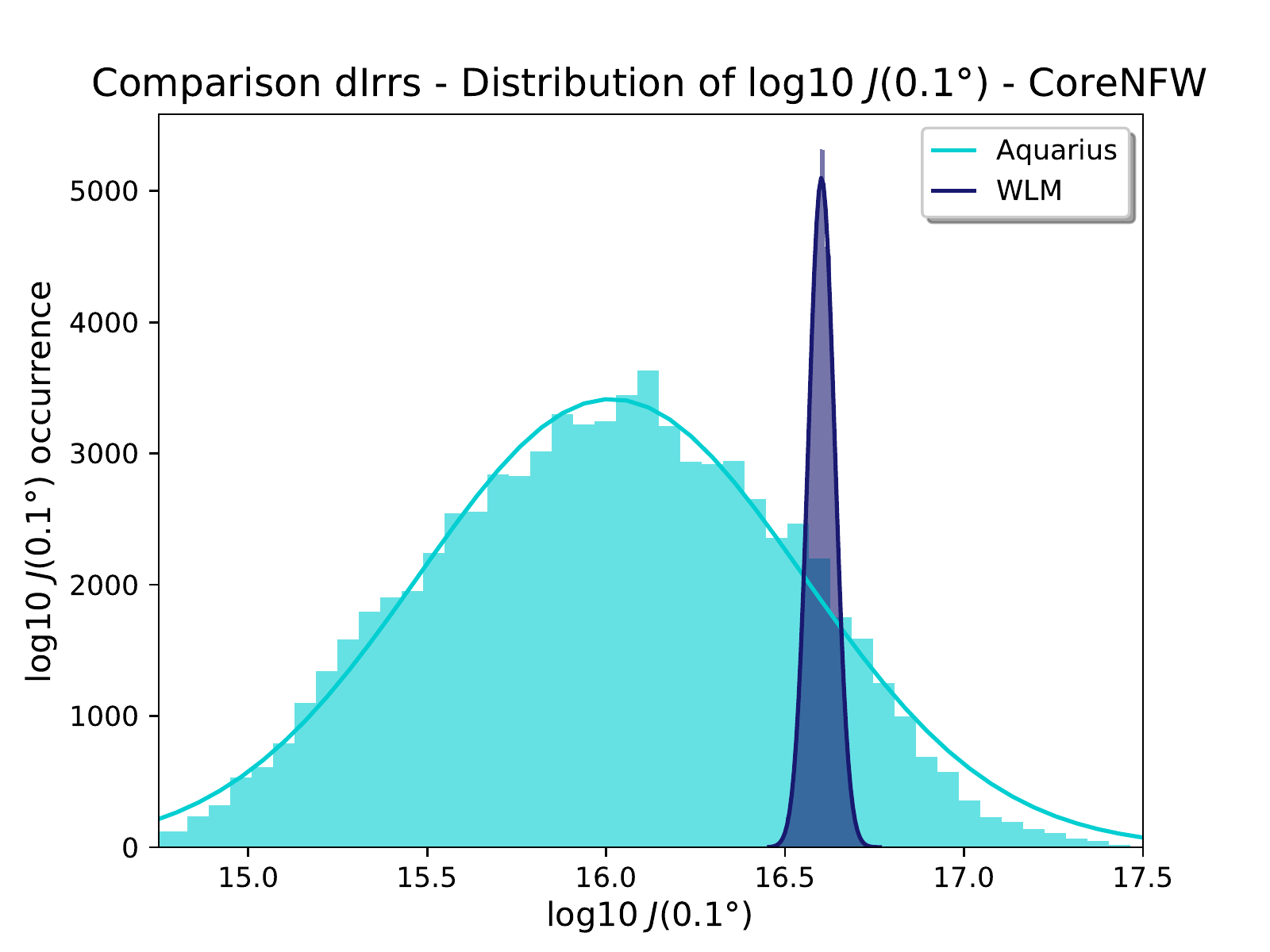}}
\caption{Histograms and fits (solid lines) of the results of the MCMC on the coreNFW profile parameters \cite{R18} for a ROI (Region of Interest) of 0.1\degree. \textbf{Left:} Distribution of the $\mathscr{J}$ factor for WLM. The nominal $\mathscr{J}$ factor and its uncertainties are the mean and $\sigma$ values respectively of the fit. \textbf{Right:} Comparison of the $\mathscr{J}$ factor of WLM and Aquarius and their uncertainties on $\mathscr{J}$. This comparison shows WLM has a larger $\mathscr{J}$ factor with smaller uncertainties than Aquarius.}
\label{J_factor_fit}
\end{figure}

This new profile takes into account the history of the stellar component within the galaxy which is still active and impacts the DM distribution. $\rho_{\mathrm{NFW}}$ is the original NFW profile, $M_{\mathrm{NFW}}$ is the mass of the galaxy at some radius $r$ and $f^n$ is responsible for generating a shallower density profile at radii $r<r_c$, with $r_c$ being the core radius and where $n$ is a coefficient tied to the total star formation time.
Fitting the results of an MCMC on the coreNFW profile parameters, we derive a $\mathscr{J}$ factor of $\log_{10} \mathscr{J} (\text{GeV}^2.\text{cm}^{-5}) = 16.6 \pm 0.037$ (Fig. \ref{J_factor_fit}) based on the DM profile derived in \cite{R18}.
WLM represents a very promising target among the dIrrs as it possesses one of the highest $\mathscr{J}$ factor with extremely small uncertainties compared to those of some other dIrrs (eg. Aquarius). 

\section{Observations and data analysis}
In 2018, H.E.S.S. collected about 19 hours of data towards WLM with an offset of 0.5\degree and 0.8\degree. We perform the analysis of this dataset in order to identify a potential signal from DM. As the signal-to-noise ratio gives a maximum at an extension of 0.08\degree, this analysis is performed over a ROI of 0.1\degree, which corresponds to the point-like source treatment in H.E.S.S. We use the Mono standard configuration which only includes the events detected by the CT5 telescope.

The analysis gives the number of $\gamma$-ray-like events detected in the ON region, where the signal is expected, and the OFF region to compute the background noise. The ON region corresponds to a disk of 0.1\degree angular radius in the direction of the source while the region OFF is defined according to the multiple-OFF method.
This method allows the estimation of the residual background and the measure in the ON region simultaneously so that both are performed in the same conditions of observation and is described in \cite{Berge}.
As the ON region and all the OFF regions combined cover a different area, the acceptance corrected exposure ratio $\alpha$ is also provided which renormalizes the OFF region to the ON region area. From the analysis, we also obtain the $\gamma$ excess and its significance $\sigma$. 
Table \ref{table_results_HESS} summarizes the results of the analysis. We can conclude from it that no significant excess in the signal region has been observed towards WLM.

\begin{table}[H]
\centering
\small{
\begin{tabular}{|c||c|c|c|c|c|c|}
 \hline 
 dIrr & $N_{\text{ON}}$ & $N_{\text{OFF}}$& $\alpha$ & Live hours & $\gamma$ excess & $\sigma$   \\
 \hline
 WLM & 1677  & 26726   & 16.24 & 18.6 &31.2  & 0.7  \\
 \hline
\end{tabular}}
\caption{Data analysis results of WLM. $N_{\text{ON}}$ and $N_{\text{OFF}}$ are the number of events detected in the ON and OFF regions, $\alpha$ is the acceptance corrected exposure ratio, the live hours give the observation time, $\gamma$ gives the excess detected and the standard deviation $\sigma$ the significance of the excess.}
\label{table_results_HESS}
\end{table}

\begin{figure}[H]
\centering{\includegraphics[scale=1]{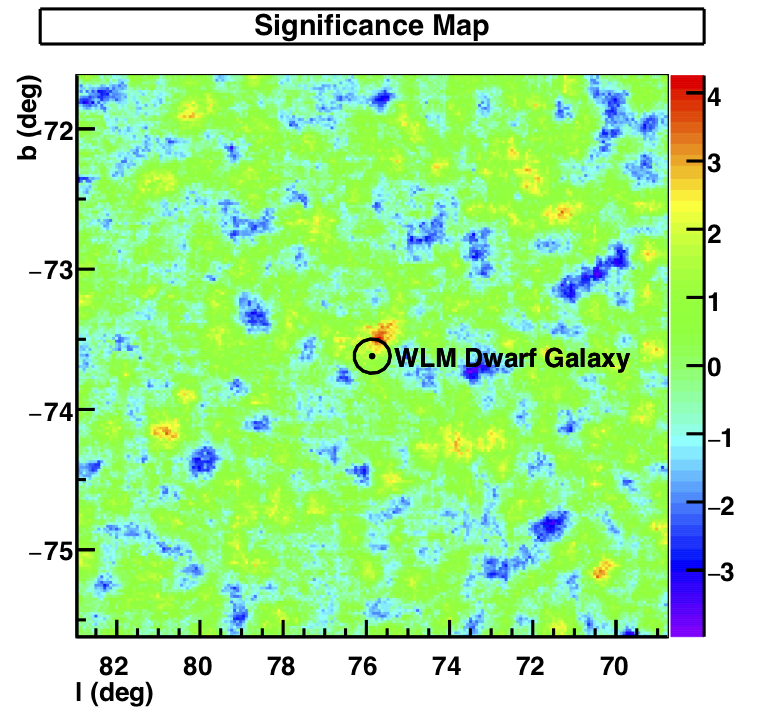}}
\caption{Significance map showing no excess in the ROI.}
\label{significance}
\end{figure}

This result can also be seen in the significance map (Fig. \ref{significance}) 
where no excess is observed in the ROI.

\section{Statistical analysis and upper limits}

A loglikelihood ratio test is performed on the data in order to constrain DM and set some upper limits on the DM annihilation cross section.


The total likelihood function contains two terms, a product of a Poisson likelihood $\mathcal{L^P}_i$ on the events of all energy bins and a log-normal distribution $\mathcal{L^J}$ of the $\mathscr{J}$ factor. This expression is written as

\begin{equation}
\mathcal{L} = \Pi^i \mathcal{L^P}_i (N_{S_i}, N_{B_i}|N_{ON}, N_{OFF},\alpha) \cdot \mathcal{L^J}(\mathscr{J}|\bar{\mathscr{J}},\sigma).
\end{equation}

For an energy bin $i$, the likelihood function $\mathcal{L^P}_i$ of the event counts is the product of two Poisson likelihoods, one for each of the ON and OFF regions:

\begin{equation}
\mathcal{L^P}_i = \frac{(N_{S_i} + N_{B_i})^{N_{\text{ON}_i}  }}{N_{\text{ON}_i}!} \exp{-(N_{S_i} + N_{B_i})} \cdot \frac{(\alpha N_{B_i})^{N_{\text{OFF}_i}}}{N_{\text{OFF}_i}!} \exp{(-\alpha N_{B_i})}
\end{equation}

where $N_{S_i}$ and $N_{B_i}$ are the number of signal events and background events respectively for a given energy bin $i$ and $\alpha$ is the ratio of the solid angles of the ON and OFF regions.

To take into account the uncertainty on the $\mathscr{J}$ factor in our analysis, we introduce a log-normal distribution in the construction of our total likelihood function $\mathcal{L}$ which is given by

\begin{equation}
\mathcal{L^J}= \frac{1}{\sqrt{2\pi} \sigma_\mathscr{J} \mathscr{J}} \exp{-\frac{( \log_{10}\mathscr{J} - \log_{10}\bar{\mathscr{J}})}{2 \sigma^2_\mathscr{J}}}.
\label{L^J}
\end{equation}

We perform a loglikelihood ratio test on the Poisson likelihood to set upper limits at 95\% C.L. on the annihilation cross section $\langle \sigma v \rangle$ based on the method \cite{cowan}.

%
If the test statistics $TS$ is less than 2.71, then the null hypothesis $\mathcal{H}_0$ is valid at 95\% C.L., whereas if $TS$ is greater 2.71, $\mathcal{H}_0$ is rejected.
This criterion is used to set the upper limits on $\langle \sigma v \rangle$.

The nuisance parameter on $\mathscr{J}$ that models its uncertainties is included afterwards in the statistical analysis using the following property \cite{FERMI_MAGIC}:

\begin{equation}
\langle \sigma v \rangle_{95\% C.L.} = \langle \sigma v \rangle_ 0 \frac{\bar{\mathscr{J}}}{\mathscr{J}_{\text{unc}}}
\end{equation}
 with $\langle \sigma v \rangle_{95\% C.L.}$ being the actual upper limits on the annihilation cross section, $\langle \sigma v \rangle_ 0$ the upper limits computed without the uncertainties on the $\mathscr{J}$ factor, $\bar{\mathscr{J}}$ the nominal or mean value of $\mathscr{J}$ and $\mathscr{J}_{\text{unc}}$ the value of the $\mathscr{J}$ factor that maximizes Eq. \eqref{L^J}.
This property allows a faster computational time for the statistical analysis.

\section{Results}

\begin{figure}[H]
\centering{\includegraphics[scale=0.4]{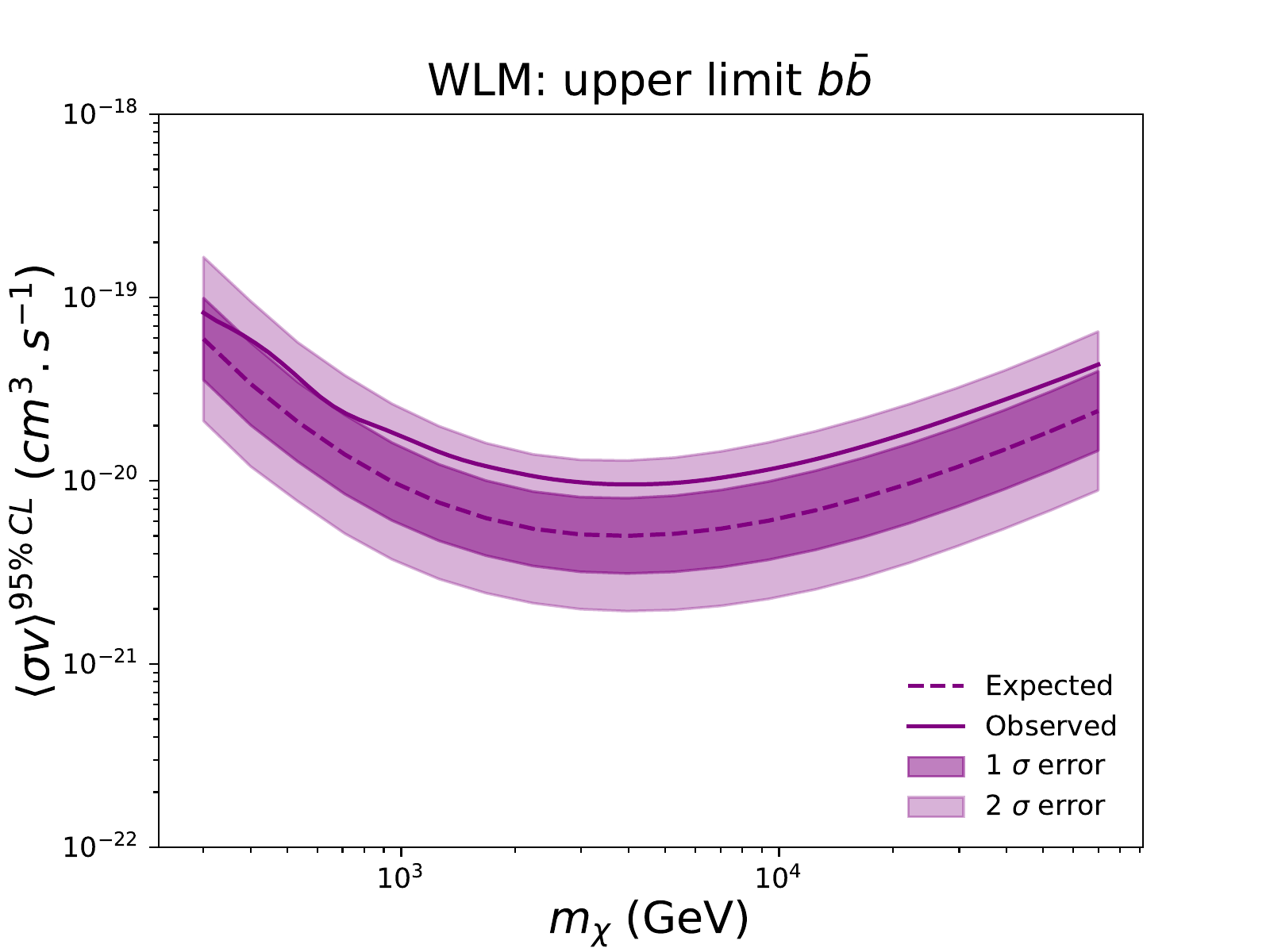} \includegraphics[scale=0.4]{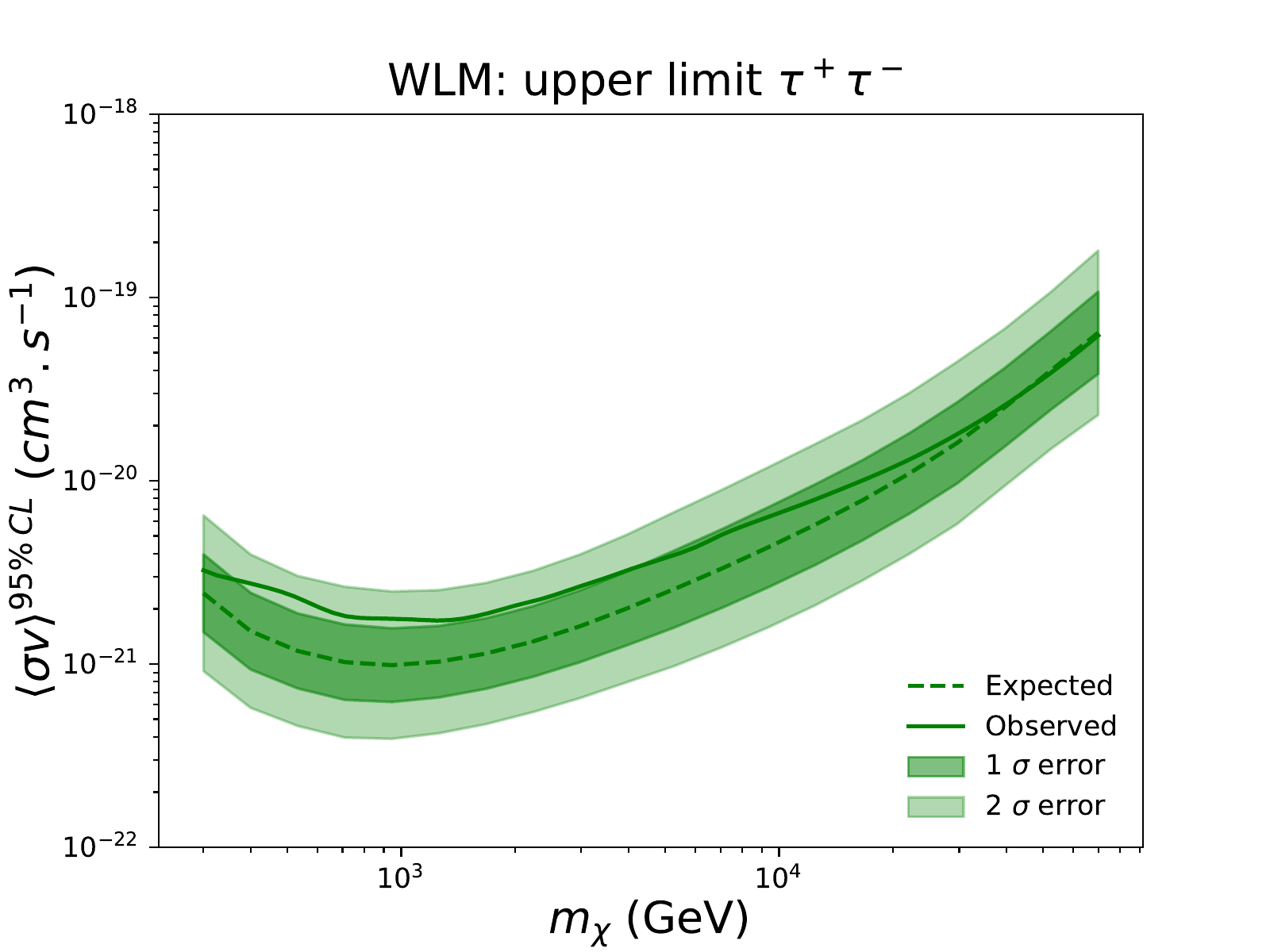}}\\
\centering{\includegraphics[scale=0.4]{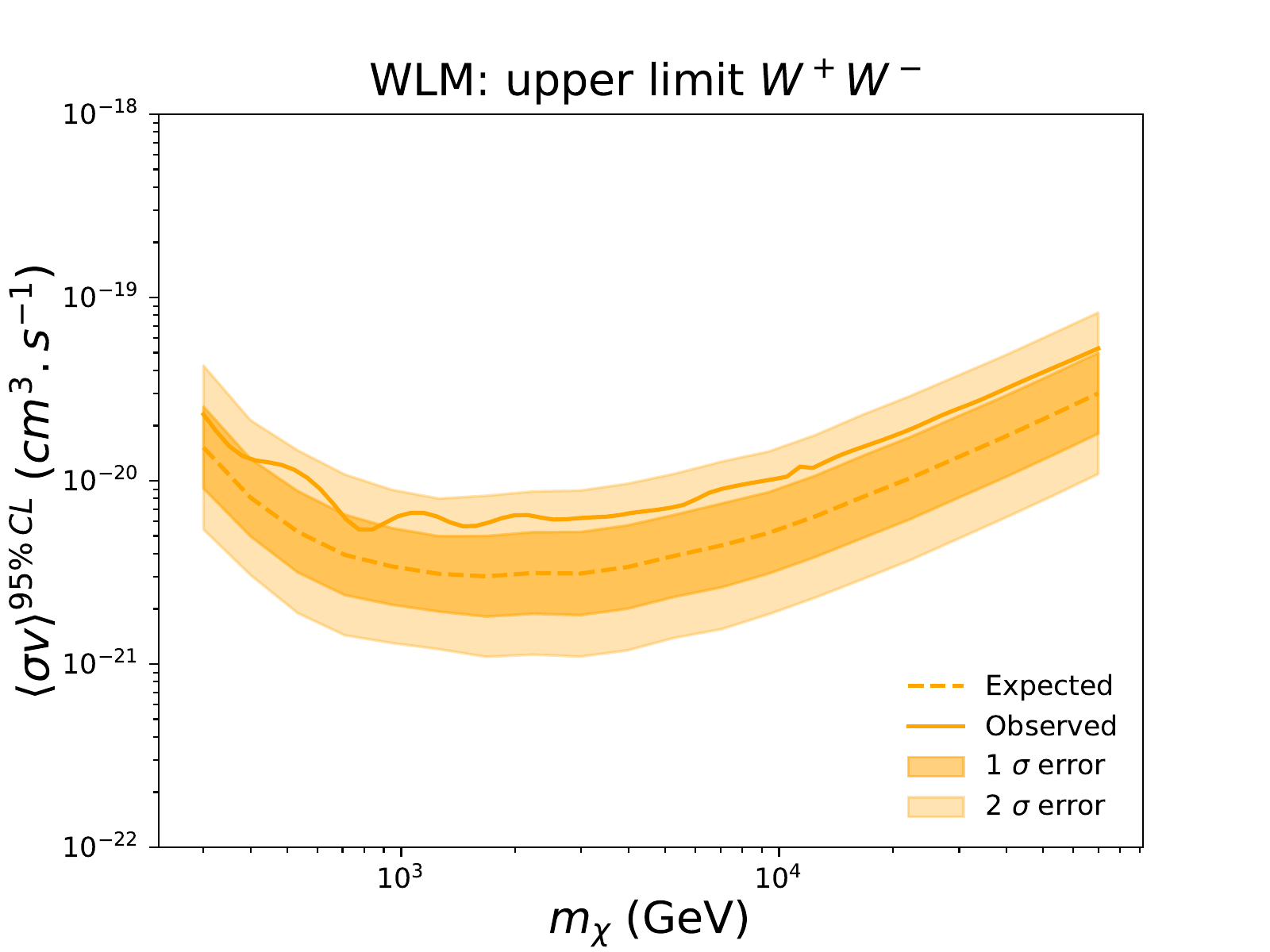} \includegraphics[scale=0.4]{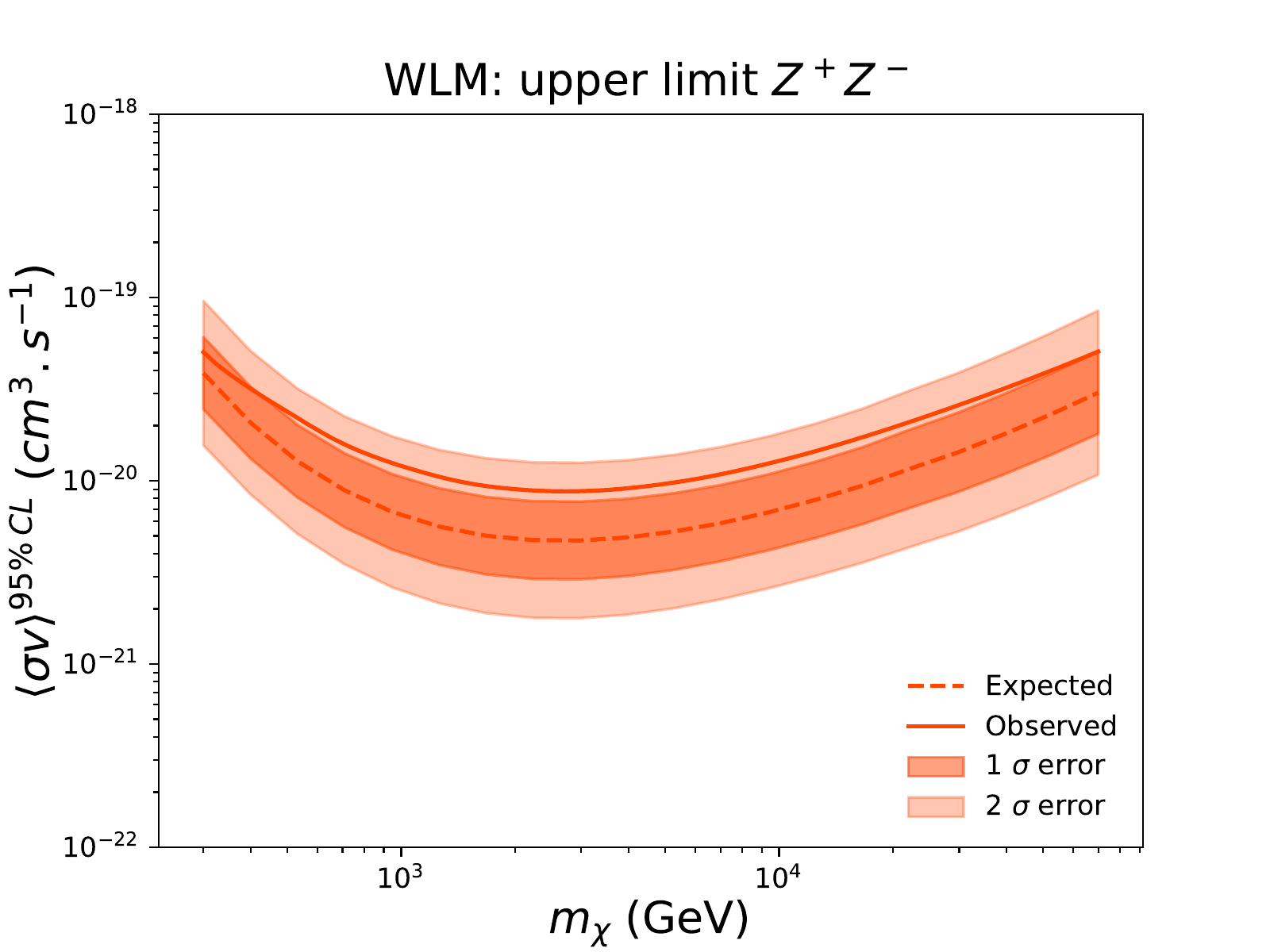} }\\
\caption{Upper limits on the annihilation cross section $\langle \sigma v \rangle$ at 95\% C.L. for WLM, in the $b\bar{b}$, $\tau^+\tau^-$, $W^+W^-$, $Z^+Z^-$ annihilation channels. These upper limits include the uncertainties on the $\mathscr{J}$ factor. The solid lines are the observed limits, the dashed lines the mean expected limits and the dark (resp. light) bands are the 1 $\sigma$ (resp. 2 $\sigma$) containment bands.}
\label{UL_sigmav}
\end{figure}

As no significant excess has been found towards WLM in the ROI, upper limits on the DM annihilation cross section $\langle \sigma v \rangle$ at 95\% C.L. vs. the DM mass are computed using the loglikelihood ratio method for the $b\bar{b}$, $\tau^+\tau^-$, $W^+W^-$ and $Z^+Z^-$ annihilation channels (fig. \ref{UL_sigmav}).
Each annihilation channel is treated individually which corresponds to a branching ratio of $B_f = 100\%$ and all the spectra are simulated using Pythia \cite{cirelli}. We also include the uncertainties on $\mathscr{J}$ as a nuisance parameter in our analysis which makes the derivation of the upper limits more conservative. Figures \ref{UL_sigmav} 
shows the upper limits obtained for all these annihilation channels with the solid lines being the observed limits, the dashed lines the mean expected limits and the dark (resp. light) bands representing the 1 $\sigma$ (resp. 2 $\sigma$) uncertainty bands. The mean expected limits and 1-2$\sigma$ containment bands are derived from a sample of 100 Poisson realizations of the background events in the ON and OFF regions. The mean expected limits corresponds to the mean of the distribution of $\log_{10} \langle \sigma v \rangle$ on these 100 Poisson realizations and the uncertainty bands are given by the standard deviation of this distribution.

The observed upper limits on $\langle \sigma v \rangle$ at 95\% C.L. reach the magnitude of $\langle \sigma v \rangle \sim 10^{-20}$ $\text{cm}^3.\text{s}^{-1}$ in the quark and boson annihilation channels at a DM mass of 1TeV.
They improve by an order of magnitude in the leptonic annihilation channel with a $\langle \sigma v \rangle \sim 10^{-21}$ $\text{cm}^3.\text{s}^{-1}$ at 1 TeV. 


\section{Conclusion}

With its recent 19 hour observations towards WLM, H.E.S.S. is the first IACT experiment to observe a dwarf irregular galaxy to search for DM annihilation signals. As no detection of a significant signal has been made in the ROI, upper limits on the annihilation cross section at 95\% C.L. have been derived for many individual annihilation channels with a branching ratio of $B_f = 100\%$. In the case of a continuum spectrum, the most constraining limits are given by the $\tau^+\tau^-$ channel with a $\langle \sigma v \rangle \sim 10^{-21}$ $\text{cm}^3.\text{s}^{-1}$ at a DM mass of 1TeV. 
The upper limits derived in this work improve of a factor of 10 to almost 100 compared to those obtained by the HAWC experiment \cite{proceedingHAWC}.

\section{Acknowledgements}

We thank Francesca Calore, LAPTh Annecy, for the useful discussions on the theoretical part of this study, as well as Justin Read, University of Surrey, for his insight about the dark matter distribution of WLM.

The support of the Namibian authorities and of the University of Namibia in facilitating the construction and operation of H.E.S.S. is gratefully acknowledged, as is the support by the German Ministry for Education and Research (BMBF), the Max Planck Society, the German Research Foundation (DFG), the Helmholtz Association, the Alexander von Humboldt Foundation, the French Ministry of Higher Education, Research and Innovation, the Centre National de la Recherche Scientifique (CNRS/IN2P3 and CNRS/INSU), the Commissariat \`a l'\'energie atomique et aux \'energies alternatives (CEA), the U.K. Science and Technology Facilities Council (STFC), the Knut and Alice Wallenberg Foundation, the National Science Centre, Poland grant no. 2016/22/M/ST9/00382, the South African Department of Science and Technology and National Research Foundation, the University of Namibia, the National Commission on Research, Science \& Technology of Namibia (NCRST), the Austrian Federal Ministry of Education, Science and Research and the Austrian Science Fund (FWF), the Australian Research Council (ARC), the Japan Society for the Promotion of Science and by the University of Amsterdam.

We appreciate the excellent work of the technical support staff in Berlin, Zeuthen, Heidelberg, Palaiseau, Paris, Saclay, T\"ubingen and in Namibia in the construction and operation of the equipment. This work benefitted from services provided by the H.E.S.S. Virtual Organisation, supported by the national resource providers of the EGI Federation.

\end{document}